\documentclass[[final,1p]{elsarticle}

\usepackage{graphics}
\usepackage{graphicx}

\usepackage{amssymb}
\usepackage{amsmath}
\usepackage{tikz}

\newcounter{bla}

\journal{Computer Physics Communications}

\begin{document}

\begin{frontmatter}



\title{COLOSS: Complex-scaled Optical and couLOmb Scattering Solver}


\author[a]{Junzhe Liu}
\author[a]{Jin Lei\corref{author}}
\author[a]{Zhongzhou Ren}

\cortext[author] {Corresponding author.\\\textit{E-mail address:} jinl@tongji.edu.cn}
\address[a]{School of Physics Science and Engineering, Tongji University, Shanghai 200092, China.}

\begin{abstract}
We introduce COLOSS, a program designed to address the scattering problem using a bound-state technique known as complex scaling. In this method, the oscillatory boundary conditions of the wave function are transformed into exponentially decaying ones, accommodating the long-range Coulomb interaction. The program implements the Woods-Saxon form of a realistic optical potential, with all potential parameters included in a well-designed input format for ease of use. This design offers users straightforward access to compute \(S\)-matrices and cross-sections of the scattering process. We provide thorough discussions on the precision of Lagrange functions and their benefits in evaluating matrix elements. Additionally, COLOSS incorporates two distinct rotation methods, making it adaptable to potentials without analytical expressions. Comparative results demonstrate that COLOSS achieves high accuracy when compared with the direct integration method, Numerov, underscoring its utility and effectiveness in scattering calculations.

\end{abstract}

\begin{keyword}
Complex scaling; Scattering theory; Nuclear reaction; Optical potential.

\end{keyword}

\end{frontmatter}



{\bf PROGRAM SUMMARY/NEW VERSION PROGRAM SUMMARY}

\begin{small}
\noindent
{\em Program Title:COLOSS}                                          \\
{\em CPC Library link to program files:} (to be added by Technical Editor) \\
{\em Developer's repository link:} https://github.com/jinleiphys/COLOSS \\
{\em Code Ocean capsule:} (to be added by Technical Editor)\\
{\em Licensing provisions(please choose one):} GPLv3  \\
{\em Programming language: Fortran}                                   \\
{\em Supplementary material:}                                 \\
{\em Journal reference of previous version:}*                  \\
{\em Does the new version supersede the previous version?:}*   \\
{\em Reasons for the new version:*}\\
{\em Summary of revisions:}*\\
{\em Nature of problem(approx. 50-250 words): The study of elastic scattering between nuclei is a fundamental problem in nuclear physics, key to understanding nuclear interactions and structure. Traditional methods for solving the Schrödinger equation in such contexts often require imposing boundary conditions at large distances, which can be computationally challenging and prone to inaccuracies, especially for reactions involving strong Coulomb interactions and complex potentials. The complex scaling method offers a robust alternative by transforming the scattered wave function from an oscillatory to an exponentially decaying form, thus eliminating the need for boundary conditions. However, implementing this method requires careful numerical handling and validation of the analytic properties of the involved potentials, such as the Woods-Saxon function, on the complex plane. Additionally, ensuring numerical stability and accuracy across different rotational techniques and integration methods is crucial. This study addresses these challenges by developing a program that leverages the complex scaling method, providing a flexible and accurate tool for calculating elastic scattering between nuclei. The program's ability to handle various optical model potentials and its validation against established methods like Numerov underscores its utility and reliability in nuclear physics research.}\\
{\em Solution method(approx. 50-250 words): To address the challenges in calculating elastic scattering between nuclei, we utilize the complex scaling method, which transforms the Schrödinger equation to simplify boundary conditions by converting the radial coordinate into the complex plane. This transformation changes the wave function from oscillatory to exponentially decaying. Our approach includes validating results against traditional methods like the Numerov algorithm to ensure accuracy. Additionally, we develop a flexible computational program capable of handling various optical model potentials, such as the Woods-Saxon potential, and performing complex scaling and integration efficiently. This method provides a robust, accurate, and computationally efficient solution for studying elastic scattering in nuclear physics.}\\
{\em Additional comments including restrictions and unusual features (approx. 50-250 words): Our method, while effective, has some restrictions and unusual features. The complex scaling method requires careful handling to avoid numerical instabilities, especially at large radial distances. Gauss-Legendre quadrature, chosen for its convergence, demands precise selection of points and weights. The computational program's flexibility to handle various optical model potentials, like the Woods-Saxon potential, adds complexity and requires thorough validation. Additionally, large-scale simulations may require significant computational resources. Despite these challenges, our approach provides precise and efficient solutions for elastic scattering in nuclear physics, though users must be cautious of potential numerical instabilities and complexities.}\\

\section{\label{sec:intro}Introduction}
Nuclear reactions play a crucial role in advancing our understanding of the synthesis of nuclei and the properties of dense matter, particularly as observed in neutron stars~\cite{FRIB}. These reactions are fundamental to various astrophysical processes, including those occurring in stellar environments and during explosive events like supernovae. By investigating rare isotopes, researchers can gain insights into the mechanisms of nucleosynthesis~\cite{RevModPhys.84.353}—the formation of new atomic nuclei from pre-existing nucleons (protons and neutrons)—and the behavior of matter under extreme conditions, such as those found in neutron stars. Understanding these reactions not only illuminates the origins of elements in the universe but also enhances our knowledge of fundamental nuclear physics and the behavior of matter at high densities.

A robust theoretical framework is essential for the accurate interpretation of nuclear reaction processes. Unlike the bound state problem, where the system's wave function exhibits an exponential decay boundary condition and thus remains localized in space, scattering problems present a primary challenge due to the non-localized nature of the wave function. This challenge is particularly pronounced when considering systems with more than two particles, where the boundary conditions become extremely complicated. In such cases, the wave function does not simply decay to zero but extends infinitely, making the problem much more difficult to handle.

Solving the scattering problem in configuration space with these complex boundary conditions requires sophisticated mathematical and numerical techniques. The non-localized nature of the wave functions in scattering problems means that they do not vanish at large distances but instead oscillate, reflecting the continuous spectrum of the system. This introduces significant difficulties in defining and implementing appropriate boundary conditions for the wave functions.

In light of these challenges, a compelling question emerges: how can bound-state-like techniques be applied to tackle scattering problems? This concept dates back to Wigner's R-matrix theory~\cite{pr.70.15,pr.70.606}, which connects bound-state and scattering problems by dividing the configuration space into an internal region with strong interactions and an external region where particles are asymptotically free. In the internal region, wave functions are treated using bound-state methods to solve the Schrödinger equation. The R-matrix, defined at the boundary, encapsulates the internal interactions and matches the internal solutions to the asymptotic solutions in the external region. This simplifies the scattering problem using bound-state-like techniques.

In addition, several other bound-state-like techniques have been developed to address the complexities of scattering problems. Notable among these are the Lorentz Integral Transform (LIT)~\cite{LAPIANA2000423}, the complex energy (CE)~\cite{prl.23.361,pra.4.1821}, and the complex scaling (CS)~\cite{PhysRevC.63.054313,PhysRevC.84.034002}. For a more comprehensive overview of these methods, readers are referred to Ref.~\cite{CARBONELL201455}. Among these techniques, the CS method is often considered the most powerful tool for directly connecting scattering with bound state problems. The CS method involves a transformation of the coordinate space into the complex plane, which effectively converts the scattering states into bound-like states. This transformation simplifies the treatment of the asymptotic boundary conditions, allowing for the use of bound-state techniques to solve scattering problems.

The CS method, initially introduced by D.R. Hartree \textit{et al}.~\cite{HMN46} for studying radio wave propagation in the atmosphere, saw a resurgence in the 1960s when Nuttal and Cohen proposed a similar approach to address scattering problems above the breakup threshold~\cite{pr.188.1542}. However, due to historical reasons, these pioneering works were interrupted. Building upon their groundwork, the complex scaling method was revisited and applied to solve resonance problems in atomic and molecular physics~\cite{HO19831,MOISEYEV1998212}, subsequently proving to be a valuable tool in quantum physics for determining the half-life of resonance states~\cite{prc.37.383,prc.71.021301}.

The method's utility quickly expanded to nuclear physics, particularly given that many exotic nuclei are unstable and exist in resonance states. The CS method is widely used to find the resonance state of cluster nuclei and determine their half-lives. While proficient at identifying bound and resonance states, the complex scaling method encounters challenges with long-range interactions, notably in scattering scenarios involving the Coulomb force~\cite{pra.12.486,pra.29.2933}, which are crucial for describing collisions with heavy nuclei. Recent advancements have adapted the method to address these long-range forces~\cite{PhysRevC.63.054313,kruppaprc}.

In this study, we present a computer code written in the Fortran language to provide a solver for the two-body scattering problem with comprehensive treatment of long-range Coulomb potentials and complex optical potentials. To make the program applicable to a wider range of potential inputs, we consider both cases where the potential has a clear analytical form and where the potential is given in a numerical form, such as those obtained from Fourier transforms in momentum space or from folding models. We have introduced two different rotation methods: rotating the basis function or rotating the Hamiltonian. The first method can be used for all cases of the potential, whereas the latter is simpler and can only be applied to potentials with analytical forms. We have developed a user-friendly input format using Fortran namelist to simplify code utilization and included three diverse examples under varying conditions to facilitate easy initiation.

The paper is organized as follows: In Sec.~\ref{sec:theory}, we provide a comprehensive overview of the theoretical framework concerning the application of the complex scaling method to quantum scattering theory and the numerical techniques involved in its implementation. Sec.~\ref{sec:program} presents a detailed description of the program, encompassing its structure, workflow, as well as input and output files. Additionally, three different examples concerning various conditions of the program are shown in Sec.~\ref{sec:example} to help users quickly start using the code. Finally, we conclude with remarks on the complex scaling method in Sec.~\ref{Conclusion}.

\section{\label{sec:theory}Theoretical framework}

\subsection{Complex Scaling Method}
The Complex Scaling (CS) method is a powerful technique used to handle continuum states in quantum mechanics. These states are characterized by wave functions that are not square integrable, meaning they do not vanish at infinity and thus cannot be normalized in the usual sense. To address this, the CS method employs a transformation that renders these exponentially divergent wave functions square integrable, making them more manageable for analysis and computation.

The key idea behind complex scaling is to rotate the coordinate along which the divergence occurs into the complex plane. This rotation reduces the oscillations in the wave functions, making them easier to handle. As a result, the wave functions become square integrable. This approach is also known as the complex coordinate method~\cite{moiseyev2011non}.

Mathematically, the complex scaling operation is represented by the transformation:
\begin{equation}
    r \rightarrow  re^{i\theta},
\end{equation}
where $ r $ is the radial coordinate and $ \theta $ is a scaling angle. This transformation rotates the coordinate $ r $ by an angle $ \theta $ in the complex plane.

The corresponding operator for this rotation transformation can be expressed as:
\begin{equation}
    \hat{S}(\theta) = e^{i\theta /2} e^{i \theta x\frac{\partial}{\partial x}},
\end{equation}
where $ \hat{S}(\theta) $ is the scaling operator. When we apply this operator to a wave function $ \Psi(r) $, we obtain:
\begin{equation}
    \hat{S}(\theta) \Psi(r) = e^{i\theta/2} \Psi(re^{i\theta}).
\end{equation}

This transformation modifies the wave function in such a way that it becomes square integrable, facilitating the study of resonance states and other continuum phenomena. By rotating the coordinates into the complex plane, the complex scaling method effectively transforms the problem into one that can be handled using standard bound state techniques.

\subsection{Basics of Scattering Theory}

Scattering theory is a fundamental framework in quantum mechanics used to describe and analyze the interactions between particles. At its core, it involves solving the Schrödinger equation for a system of two particles interacting via a potential. For two particles with a reduced mass of $\mu$ and a given angular momentum $l$, the radial Schrödinger equation in coordinate space is given by:
\begin{equation}
    H_l(r) \psi_l(r) = E \psi_l(r),
\end{equation}
where the Hamiltonian in the partial wave $l$ is defined as:
\begin{equation}
    H_l = -\frac{\hbar^2}{2\mu} \frac{d^2}{dr^2} + \frac{\hbar^2}{2\mu} \frac{l(l+1)}{r^2} + V_N + V_C,
\end{equation}
with $V_N$ representing the nuclear interaction between the projectile and the target, and $V_C$ denoting the Coulomb potential, assuming that the nuclei are uniformly charged spheres:
\begin{equation}
V_C(r) =
\begin{cases}
 (3-\frac{r^2}{R_c^2})\frac{z_1z_2e^2}{2R_c} & \text{if } r \leq R_c, \\
\frac{z_1z_2e^2}{r} & \text{if } r > R_c.
\end{cases}
\end{equation}
To handle the long-range nature of the Coulomb potential, we employ the "exterior complex scaling" method~\cite{Volkov_2009}, which separates the Coulomb interaction into long-range and short-range components:
\begin{equation}
    V_C(r) = V_C^L(r) + V_C^S(r),
\end{equation}
where the long-range Coulomb potential, arising from the interaction between two point charges, is given by:
\begin{equation}
    V_C^L(r) = \frac{z_1 z_2 e^2}{r}.
\end{equation}
The solution to the Schrödinger equation exhibits asymptotic behavior described by:
\begin{equation}
    \psi_{l}(r) \underset{r \rightarrow \infty}{\rightarrow} F_l(\eta, k r) e^{i \sigma_l} + k f_l(k) e^{i \sigma_l} O_l^{(+)}(\eta, k r),
\end{equation}
where $\sigma_l$ is the Coulomb phase shift, $\eta$ is the Sommerfeld parameter, $k$ is the wave number, $F_l$ is the regular Coulomb function, and $O_l^{(+)}(\eta, kr)$ denotes the Riccati-Hankel functions~\cite{abramowitz1972handbook} which asymptotically behave like outgoing waves:
\begin{equation}
    O_l^{(+)}(\eta, \rho) \underset{\rho \rightarrow \infty}{\longrightarrow} e^{i\left(\rho - \frac{1}{2} l \pi - \eta \ln 2 \rho + \sigma_l\right)}.
\end{equation}
A separation of the wave function can be made as:
\begin{equation}
    \psi_{l}(r) = e^{i \sigma_l} F_l(\eta, kr) + \psi_{l}^{\text{sc}}(r).
    \label{eq:OandI_seperation}
\end{equation}
In this separation, the term $F_l(\eta, kr) e^{i \sigma_l}$ represents the solution of the Schrödinger equation without any short-range interaction:
\begin{equation}
    \left[E - T_l - V_C^L \right] F_l(\eta, kr) = 0,
\end{equation}
while the second term, $\psi_{l}^{\text{sc}}(r)$, is referred to as the scattered part of the wave function.

By inserting Eq.~(\ref{eq:OandI_seperation}) into the Schrödinger equation, we obtain the inhomogeneous equation for $\psi_l^{\text{sc}}$:
\begin{equation}
    \left[E - H_l(r)\right] \psi_{l}^{\text{sc}}(r) = e^{i \sigma_l} \tilde{V}_N F_l(\eta, kr),
\end{equation}
where $\tilde{V}_N$ is defined as:
\begin{equation}
    \tilde{V}_N = V_N + V_C^S,
\end{equation}
This quantity includes only the short-range part of the potential with both Coulomb and nuclear parts, contributing to the scattering behavior of the system.

In basic scattering theory, the scattering amplitude is expressed as:
\begin{equation}
    f_l(k) = -\frac{2 \mu}{\hbar^2 k^2} e^{-i \sigma_l} \int dr F_l(\eta, kr) \tilde{V}_N(r) \psi_{l}(r).
\end{equation}
Upon separating the wave function, the scattering amplitude can be decomposed into two terms:
\begin{equation}
    f_l = f_l^{\text{Born}} + f_l^{\text{sc}},
\end{equation}
where the two terms are defined as:
\begin{equation}
    f_l^{\text{Born}}(k) = -\frac{2 \mu}{\hbar^2 k^2} \int dr F_l(\eta, kr) \tilde{V}_N(r) F_l(\eta, kr),
\end{equation}
and
\begin{equation}
    f_l^{\text{sc}}(k) = -\frac{2 \mu}{\hbar^2 k^2} e^{-i \sigma_l} \int dr F_l(\eta, kr) \tilde{V}_N(r) \psi_l^{\text{sc}}(r).
\end{equation}
The first term, $f_l^{\text{Born}}$, is referred to as the Born term, while the second term, $f_l^{\text{sc}}$, represents the scattered correction to the Born term.

The Coulomb scattering amplitude for a given scattering angle is represented by:
\begin{equation}
    f_{\mathrm{C}}(\Omega) = -\frac{\eta}{2k \sin^2 \left(\frac{1}{2} \theta\right)} e^{2i\left(\sigma_0 - \eta \ln \sin \left(\frac{1}{2} \theta\right)\right)},
\end{equation}
The scattering amplitude induced by the short-range interaction is given by:
\begin{equation}
    f_N(\Omega) = \sum_l (2l+1) e^{2i\sigma_l} f_l P_l(\cos{\theta}),
\end{equation}
where $f_l$ is the scattering amplitude for a specific angular momentum state, and $P_l(\cos{\theta})$ is the Legendre polynomial.

The differential cross section for elastic scattering is determined by the sum of the nuclear and Coulomb scattering amplitudes:
\begin{equation}
    \frac{d\sigma}{d\Omega} = \left|f_N(\Omega) + f_{\mathrm{C}}(\Omega)\right|^2.
    \label{eq:diffxsection}
\end{equation}

\subsection{Application of Complex Scaling Method to Scattering Theory}

The complex scaling method can be applied to scattering theory to handle the scattered part of the wave function more effectively. By performing the complex scaling operation on the scattered part of the wave function, we define:
\begin{equation}
\psi_{l}^{\text{sc},\theta}(r) = \hat{S}(\theta) \psi^{\text{sc}}_l(r) = e^{i\theta/2} \psi^{\text{sc}}_l(re^{i\theta}),
\end{equation}
where $\psi_{l}^{\text{sc},\theta}(r)$ is the complex scaled scattered part of the wave function. This transformation results in an exponentially decaying asymptotic behavior of the scattered part of the wave function:
\begin{equation}
\psi_{l}^{\text{sc},\theta}(r) \underset{r \rightarrow \infty}{\longrightarrow} e^{i\theta/2} k f_l(k) e^{-kr\sin{\theta}} e^{i\left(kr\cos{\theta}-\frac{1}{2}\pi l-\eta \ln 2 kre^{i\theta}+\sigma_l\right)}.
\end{equation}

It can be shown that $\psi_{l}^{\text{sc},\theta}$ satisfies the following inhomogeneous equation\footnote{Here we only assume that the potential has a local form.}:
\begin{equation}
\left(E - H_l^{\theta}(r)\right) \psi_{l}^{\text{sc},\theta}(r) = e^{i \sigma_l} e^{i\theta/2} \tilde{V}_N(re^{i\theta}) F_l\left(\eta, re^{i\theta}\right),
\label{cs_schrodinger}
\end{equation}
where the complex scaled Hamiltonian is given by:
\begin{equation}
H_l^{\theta}(r) = \hat{S}(\theta) H_l \hat{S}^{-1}(\theta) = -\frac{\hbar^2}{2\mu e^{2i \theta}} \left[\frac{\mathrm{d}^2}{\mathrm{d}r^2} - \frac{l(l+1)}{r^2}\right] + V_N(re^{i \theta}) + V_C^S(re^{i \theta}) + V_C^L(re^{i \theta}).
\end{equation}

Using the Cauchy theorem and the complex scaled scattered part of the wave function, correction of the scattering amplitude to the Born term can be expressed as:
\begin{equation}
f_l^{\text{sc}}(k) = -\frac{2\mu}{\hbar^2 k^2} e^{i \theta / 2} e^{-i \sigma_l} \int_0^{\infty} dr F_l\left(k r e^{i \theta}\right) \tilde{V}_N\left(r e^{i \theta}\right) \psi_{l}^{\text{sc}, \theta}(r).
\label{scamp_cs}
\end{equation}
To solve the complex-scaled Schrödinger equation (\ref{cs_schrodinger}), one can use a set of square-integrable basis functions $\{\phi_i\}$ to expand $\psi_{l}^{\text{sc},\theta}$ as:
\begin{equation}
\psi_{l}^{\text{sc},\theta}(r) = \sum_{i=1}^N c_i(\theta) \phi_i(r).
\label{wf_expansion}
\end{equation}
By substituting Equation (\ref{wf_expansion}) into Equation (\ref{cs_schrodinger}) and projecting onto $\phi_i$, a linear equation is obtained:
\begin{equation}
\sum_{j=1}^N \left\{E N_{ij} - H_{l,ij}^{\theta}\right\} c_j(\theta) = b_i(\theta),
\label{eq:inhomo_lineareq}
\end{equation}
where $N_{ij}=\langle \phi_i | \phi_j \rangle$ and the inhomogeneous term is:
\begin{equation}
b_i(\theta) = e^{i \theta / 2} e^{i \sigma_l} \int_0^{\infty} dr \phi_i(r) \tilde{V}_N\left(r e^{i \theta}\right) F_l\left(k r e^{i \theta}\right),
\end{equation}
and the matrix elements are given by:
\begin{equation}
H_{l,ij}^{\theta} = \int_0^{\infty} \phi_i(r) H_{l}^{\theta}(r) \phi_j(r) dr.
\label{eq:Hmat_element}
\end{equation}
To include the non-local potential, the potential matrix element should be evaluated with:
\begin{equation}
V_{ij}^\theta = \int_0^{\infty} e^{i \theta} \phi_i(r) V\left(r e^{i \theta}, r^{\prime} e^{i \theta}\right) \phi_j\left(r^{\prime}\right) dr dr^{\prime},
\end{equation}
and the inhomogeneous term is:
\begin{equation}
b_i(\theta) = e^{i \theta / 2} e^{i \sigma_l} \int_0^{\infty} dr dr' \phi_i(r) \tilde{V}_N\left(r e^{i \theta}, r' e^{i \theta} \right) F_l\left(k r e^{i \theta}\right).
\end{equation}
With the coefficients $c_j(\theta)$ obtained from solving Eq.~(\ref{eq:inhomo_lineareq}), the expansion can be inserted into Eq.~(\ref{scamp_cs}), leading to the final expression for the scattered part of the scattering amplitude:
\begin{equation}
f^{\text{sc}}_l = -\frac{2\mu}{\hbar^2 k^2} e^{-2i \sigma_l} \sum_i c_i(\theta) b_i(\theta).
\label{eq:fsc_lineareq}
\end{equation}
This method provides a systematic way to determine the scattering amplitude by expanding the complex scaled scattered wave function in terms of a basis set and solving the resulting linear equations.

Alternatively, using the Green's function method to solve the complex scaled Schrödinger equation, we can express the complex scaled scattered part of the wave function as:
\begin{equation}
\psi_{l}^{\text{sc},\theta}(r) = \int dr' G_l^\theta\left(E ; r, r^{\prime}\right) \tilde{V}_N\left(r^{\prime} e^{i \theta}\right) F_l\left(k r^{\prime} e^{i \theta}\right).
\label{wf_by_Green}
\end{equation}
The Green's function can be expanded as:
\begin{equation}
G_l^\theta\left(E ; r, r^{\prime}\right) = \sum_i \frac{\tilde{\psi}_i(r) \tilde{\psi}_i(r')}{E - \epsilon_i(\theta)},
\label{eq:Green_expand_eigen}
\end{equation}
where $\tilde{\psi}_i(r)$ are the eigenvectors of the complex scaled Hamiltonian:
\begin{equation}
H_l(\theta) \tilde{\psi}_i(r) = \epsilon_i(\theta) \tilde{\psi}_i(r).
\label{eq:eigenproblem}
\end{equation}
By combining Eq.~(\ref{scamp_cs}), (\ref{wf_by_Green}), and (\ref{eq:Green_expand_eigen}), a compact form of the scattering amplitude can be derived as:
\begin{equation}
f^{\text{sc}}_l(k) = -\frac{2\mu}{\hbar^2 k^2} e^{i\theta} \sum_{n=1}^N \frac{d_i d_i}{E - \epsilon_i(\theta)},
\label{eq:fsc_greens_expand}
\end{equation}
where $d_i$ is:
\begin{equation}
d_i = \int dr \tilde{\psi}_i(r) V_N\left(r e^{i \theta}\right) F_l\left(k r e^{i \theta}\right).
\end{equation}
This approach utilizes the Green's function to represent the complex scaled scattered wave function in terms of the eigenvectors of the complex scaled Hamiltonian, providing an alternative method to determine the scattering amplitude.

\subsection{Optical Model Potential}
In the study of nucleus-nucleus interactions, the Optical Model Potential (OMP) is frequently employed to describe the nuclear reactions. The nuclear part of the OMP often adopts a Woods-Saxon potential and its derivatives, depending on the relative coordinate $r$ between the nuclei:
\begin{equation}
V_N(r) = 
- V_0 f\left(r, R_R, a_R\right) 
- V_S \frac{d}{dr} f\left(r, R_{vs}, a_{vs}\right)
- i W_0 f\left(r, R_W, a_W\right) 
- i W_S \frac{d}{dr} f\left(r, R_{ws}, a_{ws}\right),
\end{equation}
where the Woods-Saxon function $f(r, R, a)$ is defined as:
\begin{equation}
f(r, R, a) = \frac{1}{1 + \exp\left((r - R) / a\right)}.
\end{equation}
Here, $V_0$, $W_0$, $W_S$, and $V_S$ are the potential strengths, and $R_R$, $R_W$, $R_{ws}$, and $R_{vs}$ are the corresponding radii, while $a_R$, $a_W$, $a_{ws}$, and $a_{vs}$ are the diffuseness parameters. For simplicity, here we ignore the spin-orbit coupling terms.

When extending the Woods-Saxon function into the complex plane, it exhibits a set of poles $z_n$ given by:
\begin{equation}
z_n = R + i(2n+1)\pi a, \quad n = \pm 1, \pm 2, \ldots.
\end{equation}
Figure \ref{fig:poles} illustrates these poles on the complex plane, where the red line in the first quadrant represents the integration path when rotating the potential.

To ensure the validity of the Cauchy theorem, the contour of integration should not include these poles. This requirement imposes a restriction on the rotation angle $\theta$, which should satisfy:
\begin{equation}
\tan\theta < \frac{\pi a}{R}.
\end{equation}
This restriction ensures that the contour avoids the poles of the Woods-Saxon function in the complex plane, allowing for the proper application of the complex scaling method in the calculation of the scattering amplitude with the Optical Model Potential.

By adhering to this restriction, the complex scaling method can be effectively applied to the Optical Model Potential, facilitating the analysis of scattering processes in nuclear physics. This approach ensures that the potential remains well-behaved in the complex plane, enabling accurate and reliable calculations of scattering amplitudes.

\begin{figure}
    \centering
    \includegraphics[width=0.8\textwidth]{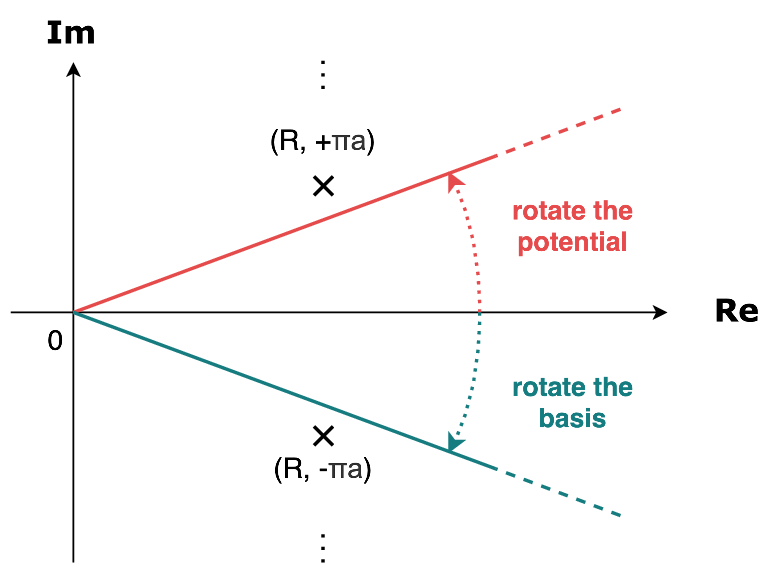}
    \caption{Diagram illustrating the poles of the Woods-Saxon function, represented by crosses. According to the Cauchy theorem, the integration contour must exclude these poles, imposing a restriction on the rotation angle.}
    \label{fig:poles}
\end{figure}

\subsection{Lagrange Functions and Lagrange-Laguerre Basis}
The Lagrange-Legendre basis is extensively employed in the $R$-matrix formalism \cite{Descouvemont_2010} within nuclear physics, which is defined in the fixed interval region $[0,1]$. This basis is particularly useful for problems where the wave function is confined to a finite interval, allowing for efficient and accurate numerical solutions. However, for the case of the current study, the wave function is defined in the region from $0$ to $\infty$, making the Lagrange-Legendre basis unsuitable for use in the complex scaling method for scattering problems. The complex scaling method requires a basis that can accurately represent functions with an infinite domain, which is not possible with the Lagrange-Legendre basis.

In our numerical implementation, we utilize Lagrange-Laguerre functions regularized by $x$ as the basis. This function, denoted by $f_j(x)$, is expressed as:
\begin{equation}
f_j(x) = (-1)^j \left(h_N^\alpha x_j\right)^{-1 / 2} \frac{L_N^\alpha(x)}{x-x_j} x^{\alpha / 2+1} e^{-x / 2},
\label{lag-lag_def}
\end{equation}
where $L_N^{\alpha}$ represents the generalized Laguerre polynomial of order $N$ with parameter $\alpha$, $h_N^\alpha$ corresponds to the square norms of the polynomial, and $x_j$ are the roots of the polynomial. The generalized Laguerre polynomials $L_N^\alpha(x)$ are orthogonal polynomials that are solutions to the Laguerre differential equation and are widely used in quantum mechanics, particularly in problems involving radial functions.

The $x$-regularization allows the selection of the $\alpha$ parameter to no longer depend on $l$~\cite{baye_lagrange-mesh_2015}. This regularization is crucial because it ensures that the basis functions have the correct asymptotic behavior at origin, which is essential for accurately representing scattering states. By decoupling the parameter $\alpha$ from the angular momentum quantum number $l$, the basis functions can be tailored to better suit the specific problem at hand, providing greater flexibility and accuracy in numerical computations.

More details regarding Lagrange functions and the Lagrange-mesh method can be found in Ref.~\cite{baye_lagrange-mesh_2015}. The Lagrange-mesh method is a powerful numerical technique that combines the efficiency of Gaussian quadrature with the flexibility of Lagrange interpolation. This method allows for the accurate and efficient computation of integrals and eigenvalue problems, making it highly suitable for a wide range of applications in quantum mechanics and other fields.

The $x$-regularized Lagrange-Laguerre functions are non-orthogonal, and the inner products between them are given by:
\begin{equation}
N_{ij} = \int f_i(x) f_j(x) \, dx = \delta_{ij} + \frac{(-1)^{i-j}}{\sqrt{x_i x_j}}.
\label{eq:innerproduc}
\end{equation}
When the order of the polynomial is large, the values of $x_i$ may extend beyond the range of the interaction, resulting in wasted computational effort during numerical integration. To enhance numerical efficiency, a scaling transformation is performed on the coordinate:
\begin{equation}
r = h_S x,
\end{equation}
where $h_S$ represents the scaling factor, and it has a unit of fm. The scaled basis is defined as:
\begin{equation}
\phi_i(r) = h_S^{-1/2} f_i(r/h_S),
\end{equation}
where the factor $h_S^{-1/2}$ is incorporated to maintain the same overlap $N_{ij}$ for the scaled basis as that of the unscaled basis. Please note that both the Lagrange-Laguerre function and variable are dimensionless. The scaling factor, which maps the Lagrange-Laguerre function and basis function, gives the unit.

Lagrange functions are advantageous for evaluating matrix elements, especially when combined with the Gauss quadrature method. The potential matrix elements of the Lagrange functions can be expressed as:
\begin{equation}
V_{ij}^{\theta} = \int dr \phi_i(r) V(re^{i\theta}) \phi_j(r) \approx V(r_ie^{i\theta}) \delta_{ij}.
\label{lag_cond}
\end{equation}
This approximation significantly simplifies the computation process by requiring only the values of the potential at the mesh points. By focusing exclusively on these specific points, it eliminates the need to account for the interactions and behaviors of the basis functions across the entire domain.

It's important to note that, although Lagrange-Laguerre functions are very convenient to use, calculating the roots of high-order polynomials remains a numerically challenging task. This is illustrated in Fig.~\ref{fig:lag_poly}, which shows the modulus of the Laguerre function of order 40 with $\alpha=0$. First, it can be observed that the modulus of the Laguerre function increases exponentially as $x$ grows. Each drop in the graph indicates the presence of a root of the Laguerre function, where $L_{40}^0(x) = 0$. This means that the Laguerre function crosses the x-axis at these points, resulting in sharp drops in the modulus. However, due to numerical inaccuracies, the value does not precisely intersect with the x-axis. Moreover, as $x$ increases, the modulus of the polynomial becomes extremely large, making the precise calculation of the roots increasingly difficult.

\begin{figure}
    \centering
    \includegraphics[width=0.8\textwidth]{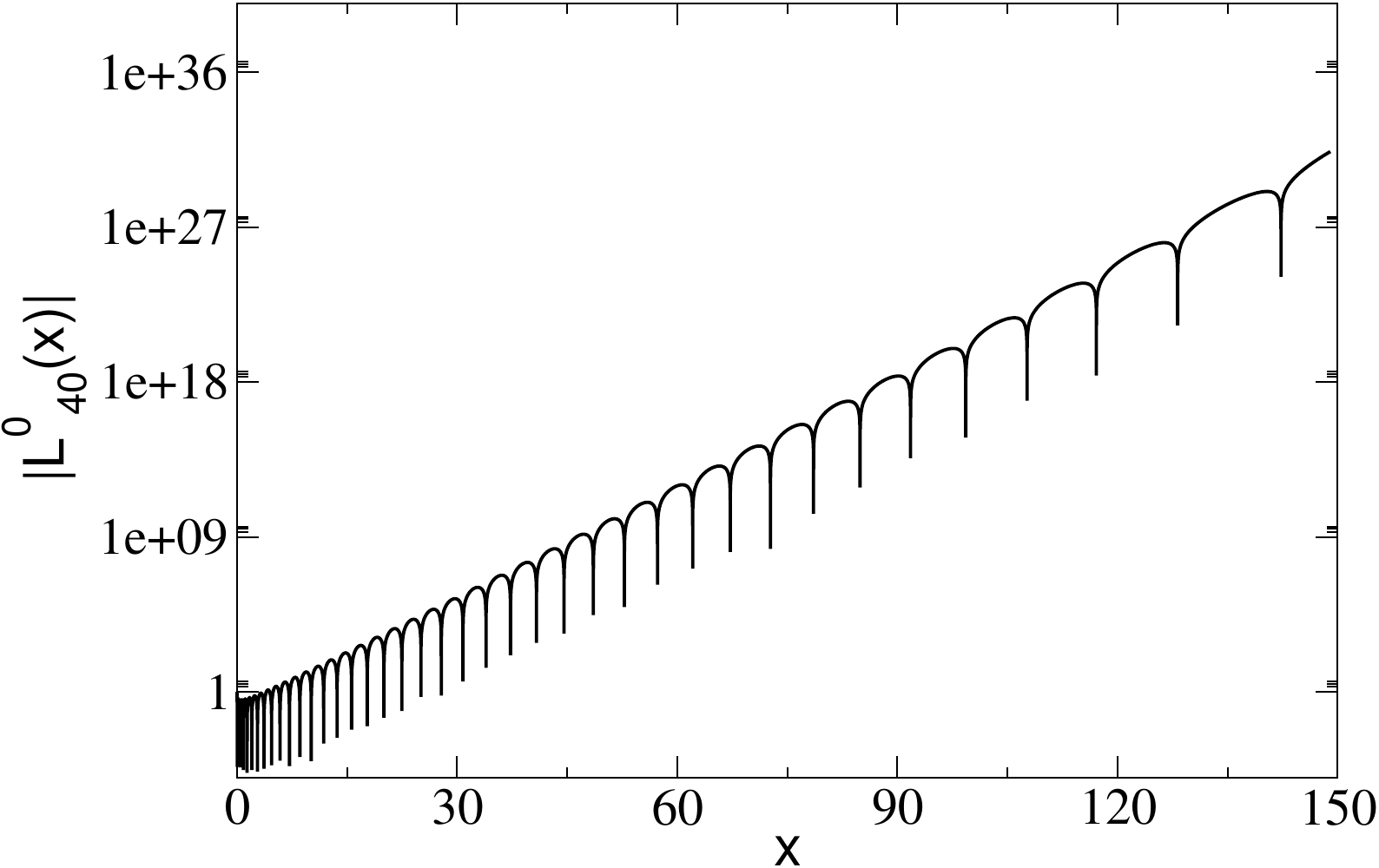}
    \caption{Absolute value of the generalized Laguerre polynomial $L_{40}^0(x)$. It is shown that the polynomial oscillates rapidly when the radius becomes very large, which makes it difficult to determine the roots of the polynomial accurately.}
    \label{fig:lag_poly}
\end{figure}

In certain scenarios where the approximation in Eq.~(\ref{lag_cond}) is not accurate enough, integration on a different mesh with more grid points may be necessary, such as on the Gauss-Legendre mesh. Instead of constructing the Lagrange-Laguerre functions from the definition provided in Eq.~(\ref{lag-lag_def}), a faster evaluation can be achieved by utilizing the roots of the Laguerre polynomial. This leads to the following expression for the basis functions:
\begin{equation}
\phi_i(x) = \lambda_i^{-1/2} \left(\frac{r}{r_i}\right)^{\alpha/2+1} e^{-(r-r_i)/2h_S} \left(\prod_{j \neq i} \frac{r-r_j}{r_i-r_j}\right),
\label{basis_inter}
\end{equation}
where $\lambda_i$ represents the weights of the Gauss-Laguerre quadrature. This alternative approach can provide a more efficient evaluation of the basis functions and matrix elements, particularly when a higher level of accuracy is required beyond the simplified approximation.

When an analytical expression for the potential is not available, such as in cases where the potential arises from a folding procedure, direct rotation of the potential function becomes challenging. To address this, a backward rotation can be introduced to the basis functions. Assuming that only discrete values for the potential are available on specific mesh points along the real axis, the integral can be transformed using the Cauchy theorem as follows:
\begin{equation}
V_{ij}(\theta) = \int_0^{\infty} dr \phi_i(r) V(re^{i\theta}) \phi_j(r) = e^{-i \theta} \int_0^{\infty} \phi_i\left(r e^{-i \theta}\right) V(r) \phi_j\left(r e^{-i \theta}\right) dr.
\label{back_rot}
\end{equation}
This transformation allows for the evaluation of the potential using values solely at particular mesh points along the real axis. It's important to note that after rotating the basis backward, the approximation in Eq.~(\ref{lag_cond}) is no longer valid. Therefore, all matrix elements must be evaluated.

Similarly, the inhomogeneous terms in Eq.~(\ref{cs_schrodinger}) can also be transformed in a similar manner:
\begin{equation}
b_i(\theta) = e^{-i \theta / 2} e^{i \sigma_l} \int_0^{\infty} dr \phi_i(r e^{-i \theta}) \tilde{V}_N\left(r \right) F_l\left(k r \right).
\end{equation}
By employing these transformations, it becomes feasible to handle situations where the potential is not analytically defined, enabling the evaluation of matrix elements and inhomogeneous terms with discrete potential values on specific mesh points along the real axis. The rotation on the basis is shown in the fourth quadrant of Fig.~\ref{fig:poles} with green lines. Although here the basis functions don't have poles, the restriction still exists to make the Cauchy theorem valid when calculating the scattering amplitude according to Eq.~(\ref{scamp_cs}).

\section{\label{sec:program}Program Description}
\subsection{Input Description}
We use the namelist feature in our programming to construct the input of the program into groups of variables, which makes it more readable and easier to use. By organizing variables into logical groups, the namelist enhances the clarity and maintainability of the code, facilitating easier debugging and modification. The following is a detailed description of the namelist:
\begin{enumerate}
\item General Namelist
\begin{itemize}
    \item \texttt{nr (integer*4)}: Number of the Lagrange-Laguerre basis used in the calculation.
    \item \texttt{alpha (real*8)}: $\alpha$ parameter of the Laguerre polynomial in Eq.~(\ref{lag-lag_def}).
    \item \texttt{Rmax (real*8)}: Maximum value of the points in the scaled Lagrange-Laguerre mesh, which satisfies:
    \begin{equation}
        R_{\text{max}} = h_s x_{\text{max}}
    \end{equation}
    \item \texttt{ctheta (real*8)}: rotation angle for complex scaling in degrees.
    \item \texttt{cwftype (integer*4)}: type of the subroutines called in the program to calculate Coulomb wave functions: 1 for \texttt{COULCC} and 2 for \texttt{cwfcomplex}.
    \item \texttt{matgauss (logical)}: boolean variable which determines whether to use the Gauss-Legendre quadrature to evaluate the matrix elements.
    \item \texttt{bgauss (logical)}: boolean variable which determines whether to use the Gauss-Legendre quadrature to evaluate the inhomogeneous terms in the linear equation.
    \item \texttt{backrot (logical)}: boolean variable which determines whether to rotate the basis backward or to rotate the potential directly in the calculation. Make sure that \texttt{matgauss} and \texttt{bgauss} are \texttt{t} before setting \texttt{backrot} as \texttt{t}.
    \item \texttt{numgauss (integer*4)}: number of Gauss-Legendre mesh points used in the evaluation of the matrix elements.
    \item \texttt{rmaxgauss (integer*4)}: maximum radius of the Gauss-Legendre mesh points.
    \item  \texttt{method (integer*4)}: option for 2 different method to calculate the scattering amplitude. Set it as 1 for linear equation method, and 2 for the Green's function method.
    \item \texttt{thetastep (real*8)}: step size for the angle in the output differential cross section.
    \item \texttt{thetamax (real*8)}: maximum value of the angle in the output differential cross section.
\end{itemize}

\item System Namelist
\begin{itemize}
    \item \texttt{zp, massp (real*8)}: charge and mass number of the projectile.
    \item \texttt{zt, masst (real*8)}: charge and mass number of the target.
    \item \texttt{jmin/jmax (integer*4)}: minimum/maximum total angular momentum of the reaction system considered in the calculation.
    \item \texttt{elab (real*8)}: incident kinetic energy of the projectile in the laboratory frame.
\end{itemize}

\item Pot Namelist
\begin{itemize}
    \item \texttt{vv, rv, av (real*8)}: depth, radius, and width parameters of the real volume term in OMP.
    \item \texttt{wv, rw, aw (real*8)}: depth, radius, and width parameters of the imaginary volume term in OMP.
    \item \texttt{vs, rvs, avs (real*8)}: depth, radius, and width parameters of the real surface term in OMP.
    \item \texttt{ws, rws, aws (real*8)}: depth, radius, and width parameters of the imaginary surface term in OMP.
    \item \texttt{rc (real*8)}: charge radius for Coulomb interaction in OMP.
\end{itemize}

\end{enumerate}

\subsection{Output Description}

\begin{enumerate}
    \item The local copy of the input file is stored in \texttt{fort.1}.
    \item The scaled Lagrange-Laguerre mesh points and weights are stored in \texttt{fort.10}.
    \item The S-matrices for different partial waves are stored in \texttt{fort.60}.
    \item The nuclear scattering amplitudes for different partial waves are stored in \texttt{fort.61}.
    \item The angular distribution of the cross section is stored in \texttt{fort.67}.
\end{enumerate}

\subsection{Structure of the Main Program and the Workflow}

The main program consists of some important subroutines, and the following is a detailed explanation of their functionalities:
\begin{itemize}
    \item \texttt{read\_input}: Reads the input variables through Fortran namelist.
    \item \texttt{init\_laguerre\_mesh}: Initializes the abscissas and weights of the Lagrange-Laguerre mesh, and carries out the scale transformation.
    \item \texttt{get\_pot\_para}: Initializes the parameters of the optical potential.
    \item \texttt{lagrange\_function}: Generates the Lagrange-Laguerre basis on the Gauss-Legendre mesh.
    \item \texttt{initial\_coul}: Generates the Coulomb wave functions on the rotated coordinate and the original coordinate.
    \item \texttt{solve\_scatt}: Generates the Hamiltonian matrix, solves the linear equation, gets the complex scaled wave function, and calculates the scattering amplitudes.
    \item \texttt{solve\_bound}: Generates the Hamiltonian matrix, solves the eigenvalue problem, and gets the eigenvalues and the eigenvectors.
    \item \texttt{solve\_scatt\_green}: Uses the eigenvalues and the eigenvectors to expand the Green's operator, and calculates the scattering amplitudes.
    \item \texttt{xsec}: Uses the scattering amplitudes to calculate the angular distribution of the differential cross section.
\end{itemize}

Figure \ref{fig:workflow} provides a detailed overview of the workflow within the primary segment of the program. The bold text within the figure denotes the names of the subroutines employed in the program, while the regular text offers concise explanations of these subroutines along with their corresponding formulas in the paper.
The left segment of the figure delineates the workflow for the linear equation method. Initially, the matrix elements of the Hamiltonian are computed according to Eq.~(\ref{eq:Hmat_element}). Subsequently, the basis's inner product is generated following Eq.~(\ref{eq:innerproduc}). Finally, the linear equation specified in Eq.~(\ref{eq:inhomo_lineareq}) is resolved, leading to the calculation of the scattered component of the scattering amplitude in accordance with Eq.~(\ref{eq:fsc_lineareq}).
Conversely, the right segment of the figure illustrates the workflow for the Green's function method. Analogous to the linear equation method, the matrix elements of the Hamiltonian and the inner product matrix are initially computed. Next, the eigenvalue problem detailed in Eq.~(\ref{eq:eigenproblem}) is solved, yielding the eigenvalues and eigenvectors. These eigenvalues and eigenvectors are then utilized to expand the Green's function through Eq.~(\ref{eq:Green_expand_eigen}), subsequently enabling the calculation of the scattered part of the scattering amplitude via Eq.~(\ref{eq:fsc_greens_expand}).
By leveraging the scattered part of the scattering amplitude derived from 2 methods, the angular distribution of the differential cross-section is finally computed following Eq.~(\ref{eq:diffxsection}).

\begin{figure}[h]
\centering
\includegraphics[width=0.6\textwidth]{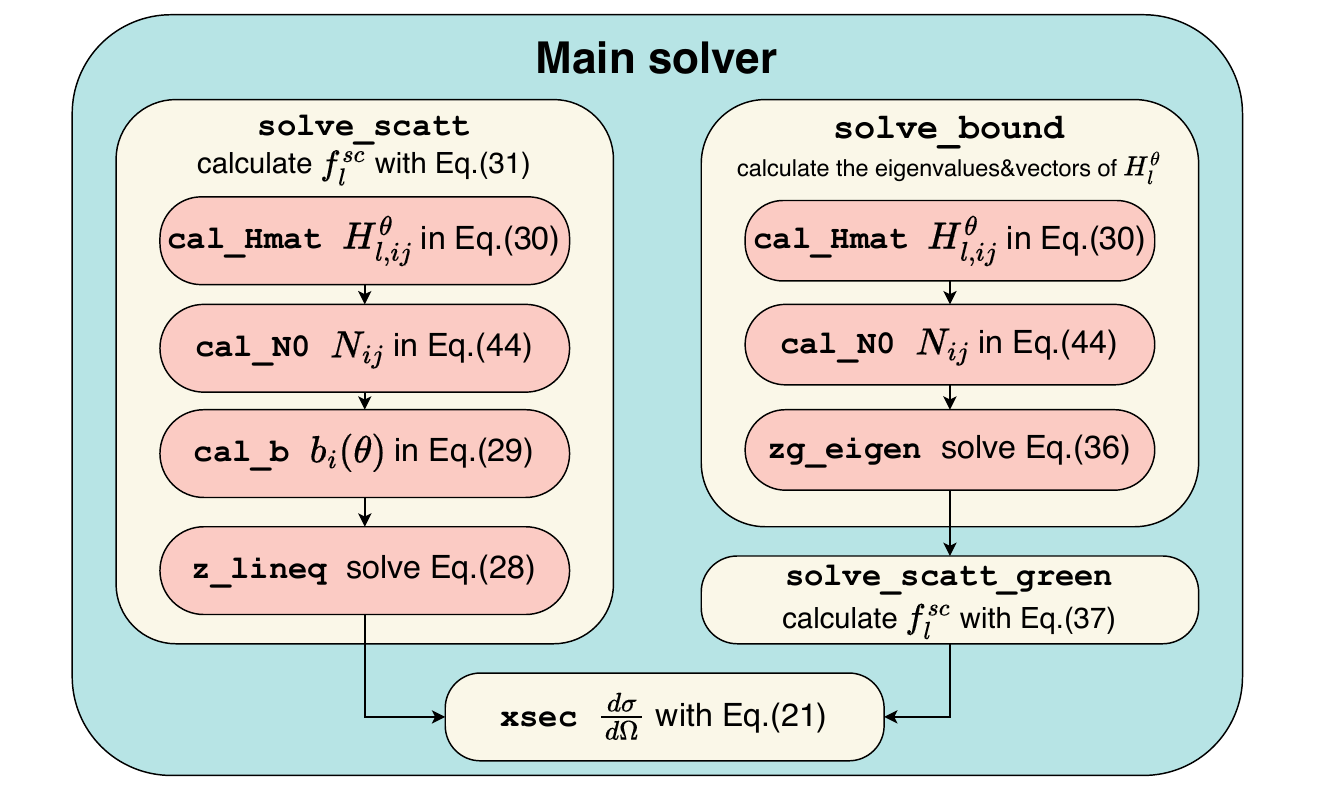}
    \caption{Workflow of the COLOSS program. See text for more details.}
    \label{fig:workflow}
\end{figure}

\subsection{List of Additional Subroutines}
In addition to the main subroutines, the program incorporates several other open-source subroutines. Below is a brief description of these subroutines:
\begin{enumerate}
    \item \texttt{COULCC}: This subroutine calculates the Coulomb wave function for complex arguments $\rho$, $\eta$, and $l$ \cite{THOMPSON1985363}.
    \item \texttt{cwfcomplex}: It is used to compute the Coulomb wave function for a broader range of complex arguments $\rho$, $\eta$, and $l$. The original code was written in C++ by N. Michel \cite{MICHEL2007232}, and an interface is provided here to allow Fortran to call it.
    \item \texttt{cdgqf}: This subroutine computes the Gauss quadrature abscissas and weights~\cite{IQPACK}. In this subroutine, it is specifically used to generate the Lagrange-Laguerre mesh.
    \item \texttt{PLM}: It calculates the associated Legendre polynomial. This subroutine is sourced from FRESCO \cite{thompson1988coupled}.
\end{enumerate}

\section{\label{sec:example}Examples}
\subsection{Installation and Compilation}
We present illustrative examples for executing the program under various conditions. A concise guide to execute the code is outlined below:

\begin{enumerate}
    \item We provide a Makefile to help compile and link all the codes. Ensure proper linkage with LAPACK by specifying the LAPACK path on your local machine in the LIB variable within the provided Makefile:
    
    \texttt{LIB = -L/path/of/your/local/lapack -llapack}
    
    \item In COLOSS, we use \texttt{gFortran} as our Fortran compiler and \texttt{GCC} to bind the C++ code with our Fortran code. Please make sure that \texttt{GCC} is installed on your machine. One can compile the program with the provided Makefile by executing:
    \texttt{> make}
    \item Transfer the executable program, \texttt{COLOSS}, to the test directory, and initiate program execution through standard input:
    
    \texttt{> ./COLOSS < inputfile}
\end{enumerate}

\subsection{Example 1: Deuteron + $^{93}$Nb Scattering}
Here we present an example to calculate the deuteron + $^{93}$Nb scattering at 20 MeV for deuteron. For the deuteron-nucleus interaction, we adopt the parameters from Ref.~\cite{PhysRevC.74.044615}. In this simple case, we take advantage of the Lagrange functions and use Eq.~(\ref{lag_cond}) to evaluate all the matrix elements and the inhomogeneous terms in Eq.~(\ref{eq:inhomo_lineareq}). Thus, the potential matrix here becomes diagonal, requiring only the values of the potential at the mesh points. We include the results of both the linear equation method and the Green's function method for comparison.

The input file for the linear equation method is:
\begin{verbatim}
&general  
    nr=60  alpha=0 Rmax=40 ctheta=6 
    matgauss=f bgauss=f method=1
    thetah=1.0 thetamax=180 /

&system 
    zp=1    massp=2   namep='2H'
    zt=41   masst=93  namet='93Nb'
    jmin=0 jmax=20  elab=20   /  

&pot
    vv=84.323 rv=1.174 av=0.809
    wv=0.351 rw=1.563 aw=0.904
    vs=0 rvs=0 avs=0
    ws=14.247 rws=1.328 aws=0.669 
    rc=1.698 /
\end{verbatim}
For the calculation with the Green's function method, change \texttt{method} to \texttt{2}.

\begin{figure}[h]
    \centering
    \includegraphics[width=0.8\textwidth]{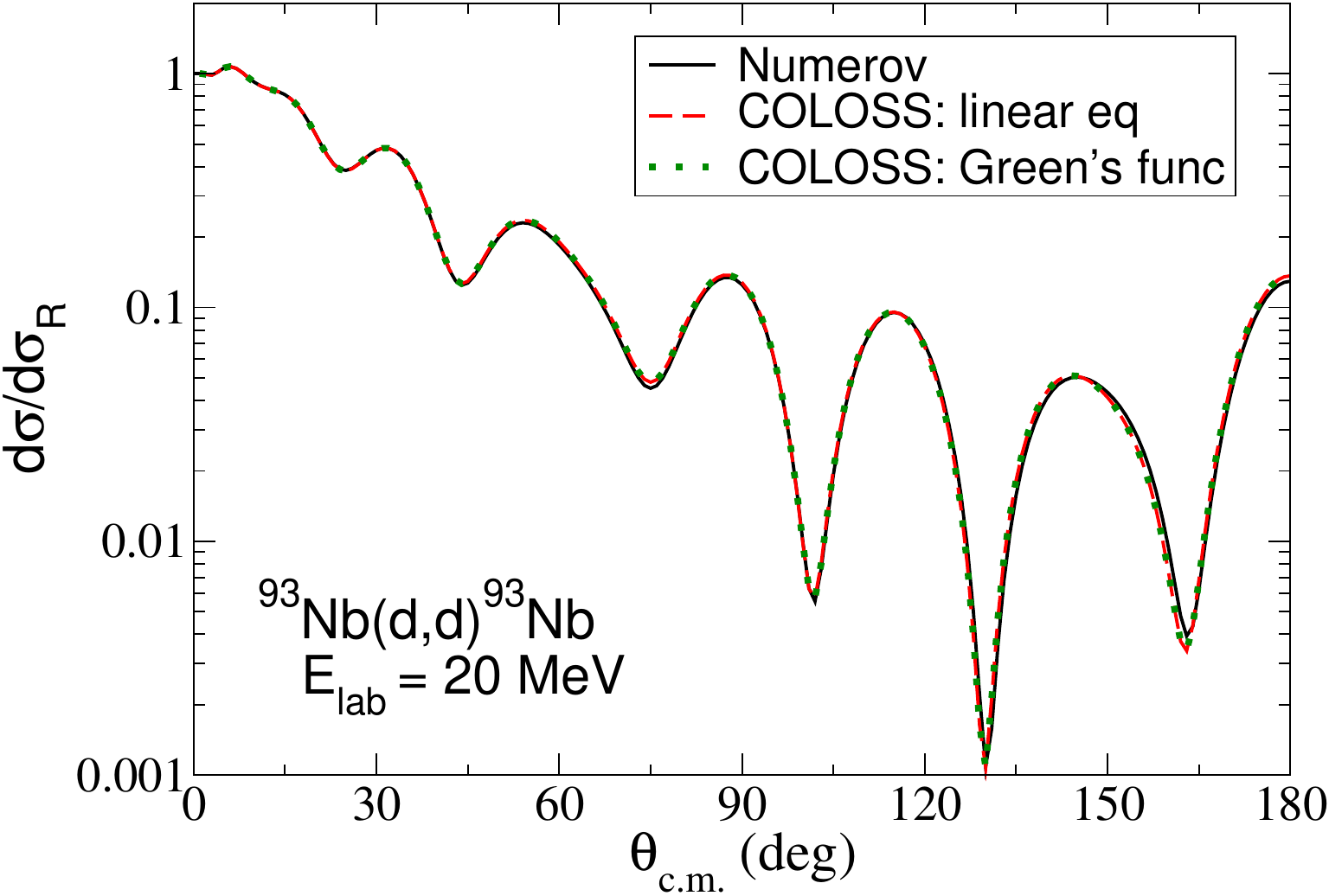}
    \caption{Cross section for deuteron + $^{93}$Nb elastic scattering at the incident energy of 20 MeV for deuteron. The black solid line represents the result from Numerov, the red dashed line represents the result given by COLOSS with the linear equation method, and the green dotted line represents the result given by COLOSS with the Green's function method.}
    \label{fig:d93Nb}
\end{figure}

The angular distribution of the cross section is illustrated in Fig.~\ref{fig:d93Nb}. In this figure, the black solid line represents the result obtained using the Numerov method, the red dashed line depicts the outcome from COLOSS using the linear equation method, and the green dotted line shows the result from COLOSS employing the Green's function method.

The results exhibit a remarkable agreement among the three methods, indicating that complex scaling can be effectively utilized with the realistic optical model potential. This concurrence also highlights the precision of the approximation described in Eq.~(\ref{lag_cond}). However, it is important to note that solving an eigenvalue problem is significantly more time-consuming compared to solving a linear equation. Consequently, the linear equation method emerges as a more practical option. The inclusion of the Green's function method serves to demonstrate the numerical equivalence between these two approaches.

\subsection{Example 2: $\alpha$+$^{28}$Si Scattering}

The second example involves $\alpha$+$^{28}$Si scattering at an incident energy of 30 MeV for the $\alpha$ particle. The $\alpha$-nucleus interaction parameters are sourced from Ref.~\cite{ch15}. In this example, both the approximation for Lagrange functions in Eq.~(\ref{lag_cond}) and the direct Gauss-Legendre quadrature method are employed to evaluate the matrix element.

The input file for the calculation using the direct Gauss-Legendre quadrature method is as follows:

\begin{verbatim}
&general  
    nr=100  alpha=0 Rmax=50 ctheta=6 
    matgauss=t bgauss=t
    numgauss=300 rmaxgauss=100
    thetah=1.0 thetamax=180 /

&system 
    zp=2    massp=4    namep='4He'
    zt=14   masst=28   namet='28Si' 
    jmin=0 jmax=20  elab=30   /  

&pot input_pot_type=1 
    vv=155.832 rv=1.342 av=0.658  
    wv=0.210 rw=1.426 aw=0.559
    vs=0 rvs=0 avs=0
    ws=25.191 rws=1.293 aws=0.636 
    rc=1.35/
\end{verbatim}

To perform the approximated calculation, one can set \texttt{matgauss} and \texttt{bgauss} to \texttt{false}. The angular distribution of the cross section is depicted in Fig.~\ref{fig:a28Si}. In this figure, the black solid line represents the result from Numerov, the red dotted line represents the result from COLOSS using Gauss-Legendre quadrature, and the green dashed line represents the result from COLOSS using the Lagrange approximation.

\begin{figure}[h]
    \centering
    \includegraphics[width=0.8\textwidth]{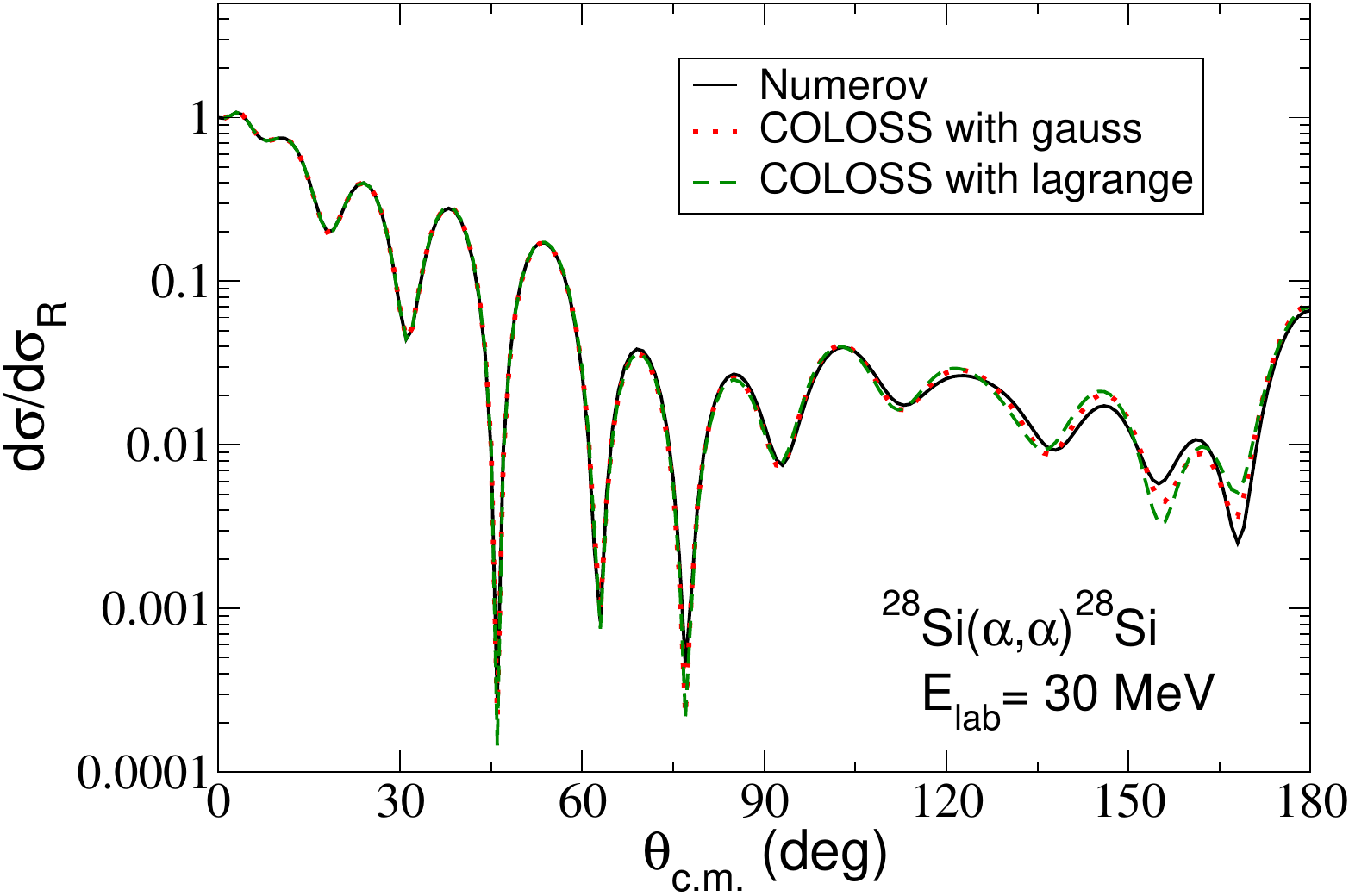}
    \caption{Cross section for $\alpha$ + $^{28}$Si elastic scattering at an incident energy of 30 MeV for the $\alpha$ particle. The black solid line represents the result from Numerov, the red dotted line represents the result from COLOSS using Gauss-Legendre quadrature, and the green dashed line represents the result from COLOSS using the Lagrange approximation.}
    \label{fig:a28Si}
\end{figure}

It is important to note that the complex-scaled Hamiltonian matrix with Lagrange-Laguerre functions is typically non-diagonal. This example aims to compare the results generated by an approximated diagonal Hamiltonian matrix with those produced by the exact non-diagonal counterpart. The figure shows that there is only a slight difference between the red dotted line and the green dashed line. Notably, for large scattering angles, the line representing the exact results is closer to the line produced by Numerov.

To compare the results of the two different integration methods in detail, we provide a table of the $S$-matrices for selected angular momenta. The results are shown in Tab.~\ref{tab:Smatrices}. The results generated with the approximation in Eq.~(\ref{lag_cond}) are labeled "Approx.", and the results generated with the direct Gauss-Legendre quadrature are labeled "Exact". For the calculation, \texttt{Rmax} is set to 50 fm, and the rotation angle is set to 6 degrees. By increasing the number of basis functions up to 100, all the converged digits are displayed in the table.

\begin{table}[h]
\centering
\begin{tabular}{c|ccc}
\hline
$S_l$  & Numerov              & Approx.             & Exact               \\ \hline
$l=11$ & (0.2230,0.0810) & (0.222,0.081) & (0.2227,0.0808) \\
$l=12$ & (0.4020,0.2203) & (0.401,0.220) & (0.4019,0.2203) \\
$l=13$ & (0.6897,0.2595) & (0.689,0.259) & (0.6897,0.2596) \\
$l=14$ & (0.8773,0.1713) & (0.8773,0.1713) & (0.8773,0.1713) \\
$l=15$ & (0.9512,0.0907) & (0.9512,0.0907) & (0.9512,0.0907) \\
\hline
\end{tabular}
\caption{Values of $S$-matrices for different angular momenta with different integration methods. The results of Numerov are included for comparison.}
\label{tab:Smatrices}
\end{table}

Upon close examination of the results, it becomes evident that the discrepancies among the three outcomes are minimal. However, a notable distinction lies in the slower convergence rate observed for the approximation method. To elucidate this phenomenon, we provide a detailed comparison focusing on the convergence behavior of the two integration methods. This analysis is systematically presented in Tab.~\ref{tab:ConvergenceComparison}. Specifically, we examine $S_{l=11}$ across varying numbers of basis functions.

\begin{table}[h]
\centering
\begin{tabular}{c|cccc}
\hline 
$S_{11}$ & NR=40           & NR=60           & NR=80           & NR=100 \\
\hline
Approx.  & (0.2544,0.1053) & (0.2235,0.0810) & (0.2224,0.0813) & (0.2225,0.0810)            \\
Exact    & (0.2237,0.0804) & (0.2227,0.0807) & (0.2227,0.0807) & (0.2227,0.0807)           \\
\hline
\end{tabular}
\caption{Convergence test of the approximation for Lagrange functions.}
\label{tab:ConvergenceComparison}
\end{table}

The comparative analysis distinctly demonstrates that the Gauss-Legendre quadrature method achieves convergence at a significantly faster rate than the approximation method. The underlying reason for this discrepancy can be attributed to the challenges associated with accurately calculating the roots of Laguerre polynomials, as depicted in Fig.~\ref{fig:lag_poly}. As previously discussed, increasing the number of basis functions inherently introduces new errors when employing Eq.~(\ref{lag_cond}) for matrix element evaluation. Conversely, the Gauss-Legendre quadrature method does not encounter such issues, owing to its inherent numerical stability and accuracy. Given these observations, it is advisable to utilize the Gauss-Legendre quadrature over the approximation method described in Eq.~(\ref{lag_cond}), particularly when dealing with a large number of basis functions.

\subsection{Example 3: $^{6}$Li + $^{208}$Pb Scattering}

The third example involves $^{6}$Li + $^{208}$Pb scattering at an incident energy of 40 MeV for $^{6}$Li. This reaction, which includes a heavy nucleus with strong Coulomb interaction, is considered at near-barrier energy. When the Sommerfeld parameter is very large, the more recently developed code by N. Michel offers enhanced numerical stability for calculating the Coulomb functions. Therefore, in this example, the \texttt{cwfcomplex} code is utilized to compute all Coulomb wave functions through an interface between Fortran and C++. Additionally, we present two different rotation methods mentioned in the previous chapter: the first is to rotate the interaction, and the second is to rotate the basis, as shown in Eq.~(\ref{back_rot}). This allows us to confirm the numerical self-consistency for different rotation methods and provide solutions when an analytical expression for the potential is not available.

The input file for rotating the potential is as follows:

\begin{verbatim}
&general  
    nr=100  alpha=0 Rmax=40 ctheta=4 
    matgauss=t bgauss=t
    numgauss = 400 rmaxgauss=150 
    thetah=0.5 thetamax=180
    cwftype=2 backrot=f/

&system 
    zp=3    massp=6     namep='6Li'
    zt=82   masst=208   namet='208Pb'
    jmin=0 jmax=40  elab=40 /  

&pot input_pot_type=1
    vv=109.500 rv=1.326 av=0.811
    wv=22.384 rw=1.534 aw=0.884
    vs=0 rvs=0 avs=0
    ws=0 rws=0 aws=0 
    rc=1.3 /
\end{verbatim}

To rotate the basis function, set \texttt{backrot} to \texttt{true}. It should be noted that, since the approximation in Eq.~(\ref{lag_cond}) cannot be applied to the rotated basis, both \texttt{matgauss} and \texttt{bgauss} must be set to \texttt{true} before setting \texttt{backrot} to \texttt{true}. The angular distribution of the cross section is shown in Fig.~\ref{fig:6Li208Pb}. In this figure, the black solid line represents the result from Numerov, the red dashed line represents the result given by COLOSS with rotation on the potential, and the green dashed line represents the result given by COLOSS with rotation on the basis. According to the figure, these three lines almost overlap, confirming the numerical self-consistency of the two different rotation methods and the accuracy of these methods compared to Numerov.

\begin{figure}[h]
    \centering
    \includegraphics[width=0.8\textwidth]{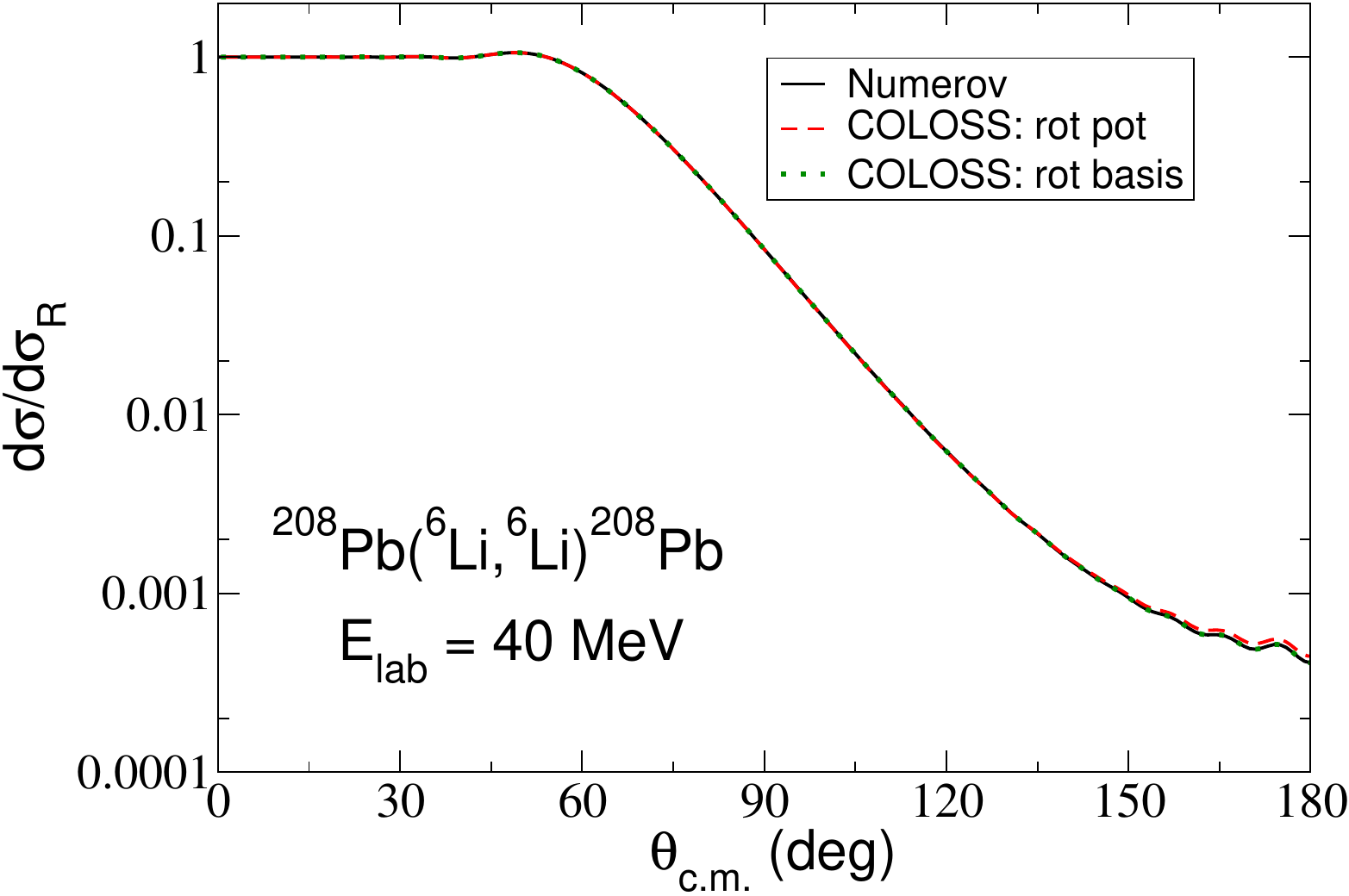}
    \caption{Cross section for $^{6}$Li + $^{208}$Pb elastic scattering at an incident energy of 40 MeV for $^{6}$Li. The black solid line represents the result from Numerov, the red dashed line represents the result given by COLOSS with rotation on the potential, and the green dashed line represents the result given by COLOSS with rotation on the basis.}
    \label{fig:6Li208Pb}
\end{figure}

\section{\label{Conclusion}Conclusion}
In this study, we introduced a program designed to calculate elastic scattering between two nuclei using the optical potential, leveraging the complex scaling method. Our program features a user-friendly input format, enabling users to adjust and utilize various parameters of the optical model potential. By applying the complex scaling method, the scattered part of the wave function transitions from an oscillatory form to an exponentially decaying one. This transformation in the asymptotic behavior obviates the necessity of imposing boundary conditions in the solution, allowing us to expand the wave function using a complete set of square-integrable basis functions.

In evaluating the matrix element of the complex-scaled Hamiltonian, we began by conducting a thorough analysis of the analytic properties of the Woods-Saxon function on the complex plane. We clarified the constraints imposed on the rotation angle by the Cauchy theorem. Subsequently, we introduced two distinct methods for numerical integration along with two rotation techniques. These technical discussions aim to enhance our understanding of the validity of the complex scaling method and the appropriateness of the chosen basis functions.

Finally, we demonstrated the program through three examples that showcase its functionality under various conditions. Each example emphasized a particular numerical aspect discussed in the paper, and we provided results obtained by Numerov for comparative analysis. The outcomes of these examples demonstrated robust agreement with Numerov's results, underscoring the program's high accuracy, the effectiveness of the complex scaling method for the optical model potential, and the numerical consistency of the program. Consequently, this program emerges as a valuable tool for applying the complex scaling method in continuum states calculations, particularly for computing elastic scattering between two nuclei.

The development of this program opens up the possibility to study more complicated problems, such as solving the inclusive breakup within the framework of quantum three-body problems. This extension would further enhance the program's applicability and contribute significantly to the field of nuclear physics.

\section*{Acknowledgements}
We are grateful to Rimantas Lazauskas for his guidance in implementing the complex scaling method. This work has been supported by the National Natural Science Foundation of China (Grants No. 12105204 and No. 12035011), by the National Key R\&D Program of China (Contract No. 2023YFA1606503), and by the Fundamental Research Funds for the Central Universities.





\end{small}

\bibliographystyle{elsarticle-num}
\bibliography{COLOSS.bib}







\end{document}